\documentclass[prb,twocolumn]{revtex4}
\usepackage{amsfonts}
\usepackage{amsmath}
\usepackage{psfrag}
\usepackage{amssymb,amsmath,amsthm}
\usepackage[dvips]{graphicx}
\usepackage{verbatim} %
\usepackage{color}
\newcommand{\beq}{\begin{equation}}
\newcommand{\eeq}{\end{equation}}
\newcommand{\bea}{\begin{eqnarray}}
\newcommand{\eea}{\end{eqnarray}}
\usepackage{color}

\newcommand{\bma}{\left(\begin{matrix}}
\newcommand{\ema}{\end{matrix}\right)}
\newcommand{\mM}{{\mathsf M}}

\def\build#1_#2{\mathrel{\mathop{#1}\limits_{#2}}}
\definecolor{pink}{rgb}{1,0.5,0.5}
\definecolor{violet}{rgb}{1,0,1} 
\definecolor{red}{rgb}{1,0,0}
\definecolor{yellow}{rgb}{0.7,1,0}
\definecolor{orange}{rgb}{1,0.5,0}
\definecolor{white}{rgb}{1,1,1}
\definecolor{blue}{rgb}{0,0,1}
\definecolor{cyan}{rgb}{0,1,1}

\begin{document}

\begin{abstract}
The coherent propagation of elastic waves in a solid filled with a random distribution of pinned dislocation segments is studied to all orders in perturbation theory. It is shown that, within the independent scattering approximation, the perturbation series  that generates the mass operator is a geometric series that can thus be formally summed. A divergent quantity is shown to be renormalizable to zero at low frequencies. At higher frequencies said quantity can be expressed in terms of a cut-off with dimensions of length, related to the dislocation length, and physical quantities can be computed in terms of two parameters, to be determined by experiment. The approach used in this problem is compared and contrasted with the scattering of de Broglie waves by delta-function potentials as described by the Schr\"odinger equation.
\end{abstract}

\title{Multiple scattering of elastic waves by pinned dislocation segments in a continuum}

\author{Dmitry Churochkin$^1$, Felipe Barra$^1$, Fernando Lund$^1$, Agnes Maurel$^2$ and Vincent Pagneux$^3$}

\affiliation{$^1$Departamento de F\'\i sica and CIMAT, Facultad de Ciencias
F\'\i sicas y Matem\'aticas, Universidad de Chile, Santiago, Chile \\
$^2$Institut Langevin, ESPCI, 1 rue Jussieu, Paris 75005, France \\
$^3$Laboratoire d'Acoustique de l'Universit\'e du Maine, \\
UMR CNRS 6613, Av. O. Messiaen, 72085 Le Mans, France}


\maketitle

\section{Introduction}
In recent years, the interaction of an elastic wave with a dislocation has been studied within the context of continuum elasticity\cite{singledisloc} (hereafter I). These results have been further used to obtain results, using multiple scattering theory, for the coherent propagation of elastic waves in the presence of many, randomly distributed, dislocations \cite{manydislocs} (hereafter II). Leading order perturbation theory was used to obtain formulae relating change in the speed of wave propagation to dislocation density. These formulae have been used to show that ultrasound---more specifically resonant ultrasound spectroscopy (RUS)---can be used as a quantitative probe of dislocation density in aluminum \cite{actamat}, with comparative advantages over X-ray diffraction (XRD) and transmission electron microscopy (TEM)\cite{JOM}.

The advent of a new, quantitative, nonintrusive, tool to probe dislocation density raises hope for progress in solving  long-standing challenges involving the plastic behavior of materials, such as fatigue \cite{fatigue}, or the brittle-to-ductile\cite{bdt} transition. More specifically, there are a number of applications that stand to benefit from a nonintrusive measurement of dislocation density. For example the plastic deformation in torsion of ice single crystals has been studied using hard X-ray diffraction \cite{ice1} in order to ascertain the role of geometrically necesary dislocations, and size effects have been unraveled through creep measurements\cite{ice2}. Also, high resolution extensometry experiments carried out on copper single crystals in tension have uncovered a rich spatiotemporal structure\cite{cu}. Distinct scales of plastic processes have been found in austenitic FeMnC steel, also using extensometry\cite{steel}. Finally, the Portevin-Le Chatelier effect\cite{plcreview}---in which plastic instabilities generate localized bands---seriously hampers the use of some alloys\cite{almg} and remains, to a large extent, a mystery. Recent attempts at developing a conceptual framewok to understand the formation of patterns by dislocations include those of Sethna and collaborators\cite{sethna}, Limkumnerd and van der Giessen\cite{vdgiessen}, and Rickman, Haataja and Le Sar\cite{lesar}.

Ultrasonic waves penetrate deep into a material and appear to be the ideal tool to probe the effects described in the previous paragraph. However, in order to probe structure, the probing wavelength must be comparable to the structure length scale. The results reported by Mujica et al. \cite{actamat} are an average over a whole sample, and rely on leading order results in a long wavelength approximation of the theory\cite{singledisloc,manydislocs}. Although these leading order results, obtained using perturbation theory in a multiple scattering framework, do provide precise and useful formulas, higher order approximations will be needed to probe shorter length scales. But, as discussed below, higher order results can diverge at high frequency because of the zero thickness of the strings used to model dislocations. This paper is devoted to the elucidation of this state of affairs.

\section{Interaction of elastic waves with dislocation segments.}
We shall use the notation of I and II: An homogeneous and isotropic, linearly elastic, medium of mass density $\rho$ is described by displacements $\vec u (\vec x,t)$ away from an equilibrium position $\vec x$ at time $t$.  In the absence of dislocations the dynamics is governed by the wave equation
\begin{equation}
\rho \frac{\partial^2 u_i}{\partial t^2}
-c_{ijkl}\frac{\partial^2 u_k}{\partial x_j\partial x_l} = 0
\label{EQonde}
\end{equation}
where the elastic constants tensor is given by $c_{ijkl}=\lambda \delta_{ij}\delta_{kl}+ \mu ( \delta_{ik}\delta_{jl}+\delta_{il} \delta_{jk} )$ with $(\lambda,\mu)$ the Lam\'e constants. Longitudinal ($L$) and transverse ($T$) waves propagate with speeds $c_L\equiv \sqrt{(\lambda+2\mu)/\rho}$ and $c_T\equiv\sqrt{\mu/\rho}$, respectively. We shall call their ratio $\gamma \equiv c_L/c_T  >1$.

A pinned dislocation segment is described by its position $\vec X(s ,t)$ as a function of a Lagrangian parameter $s$ and time, with orientation provided by the unit tangent $\hat \tau \equiv  \vec X' /|\vec X'|$, where a prime denotes derivation with respect to $s$. We shall consider unbiased edge dislocations of length $L$ and Burgers vector $\vec b$, that is, their position at equilibrium are straight lines. In the absence of external loading the dislocation dynamics is given by a linear string model
\begin{equation}
m \ddot{X}_k(s ,t) + B\dot{X}_k(s ,t) - \Gamma X_k''(s ,t)= 0
\label{eqMouv}
\end{equation}
where overdots denote derivation with respect to time, and the associated boundary conditions of pinned ends are $X_k(\pm L/2,t)=0$. In
Eqn. (\ref{eqMouv})  the coefficient
\begin{equation}
m \equiv \frac{\rho b^2}{4\pi}(1+\gamma^{-4}) \ln(\delta/\delta_0),
\label{defmasse}\end{equation}
defines a mass per unit length (with $\delta$ and $\delta_0$  the long- and  short-distance cut-off lengths, respectively),
\begin{equation}
\Gamma \equiv
\frac{\mu b^2}{2\pi} (1-\gamma^{-2})  \ln(\delta/\delta_0)
\label{deflinetension}
\end{equation}
is a line tension, and $B$ is a phenomenological viscous drag coefficient.

When a wave propagates in the presence of dislocations, there is an interaction. The behavior of the wave in this case is given, in the frequency domain, by the following equation \cite{manydislocs}  for the velocity $v_i (\vec x,\omega)$: 
\begin{equation}
-\rho \omega^2v_i(\vec x,\omega)-c_{ijkl}\frac{\partial^2}{\partial x_j\partial x_l} v_k
(\vec x,\omega)= V_{ik} v_k (\vec x,\omega)
\label{EQondes}
\end{equation}
where
\begin{equation}
\left. V_{ik}=  {\cal A} \; \mM_{ij}
 \frac{\partial}{\partial x_j}  \delta ( \vec x-\vec X_0 )\;
{\mathsf  M}_{lk}{\frac{\partial}{\partial x_l}} \right|_{\vec x=\vec X_0} \, ,
\label{potential}
\end{equation}
\bea
  {\cal A} & \equiv & \frac{8}{\pi^2}\frac{(\mu b)^2}{m} \frac{S(\omega)}{\omega^2} L \, , \\
\mM_{ij} & \equiv & t_i n_j + t_j n_i \, ,
\eea
with $\hat n \equiv \hat \tau \times \hat t$ and $\hat t \equiv \vec b/|\vec b|$ is the unit Burgers vector that indicates the direction of glide. In addition,
\beq
\frac{S(\omega)}{\omega^2} \equiv \sum_{\mbox{$N$ odd}} \frac{1}{N^2 \left( \omega^2 - \omega^2_N + i\omega B/m\right)}
\eeq
where
\beq
\omega_N \equiv \frac{N\pi}{L}\sqrt{\frac{\Gamma}{m}}
\eeq
are the eigenfrequencies of a vibrating string of length $L$, mass density $m$ and line tension $\Gamma$. Only glide motion, that is, along $\hat t$, is allowed. The potential (\ref{potential}) is a simplified expression that captures the dislocation dynamics for frequencies smaller than the first resonance frequency $\omega_1$, thus accordingly we shall use the approximation
\beq
 \frac{S(\omega)}{\omega^2} \approx \frac{1}{ \left( \omega^2 - \omega^2_1 + i\omega B/m\right)} \, .
 \label{lowS}
\eeq

\begin{figure}[h!]
\includegraphics[width=.8\columnwidth]{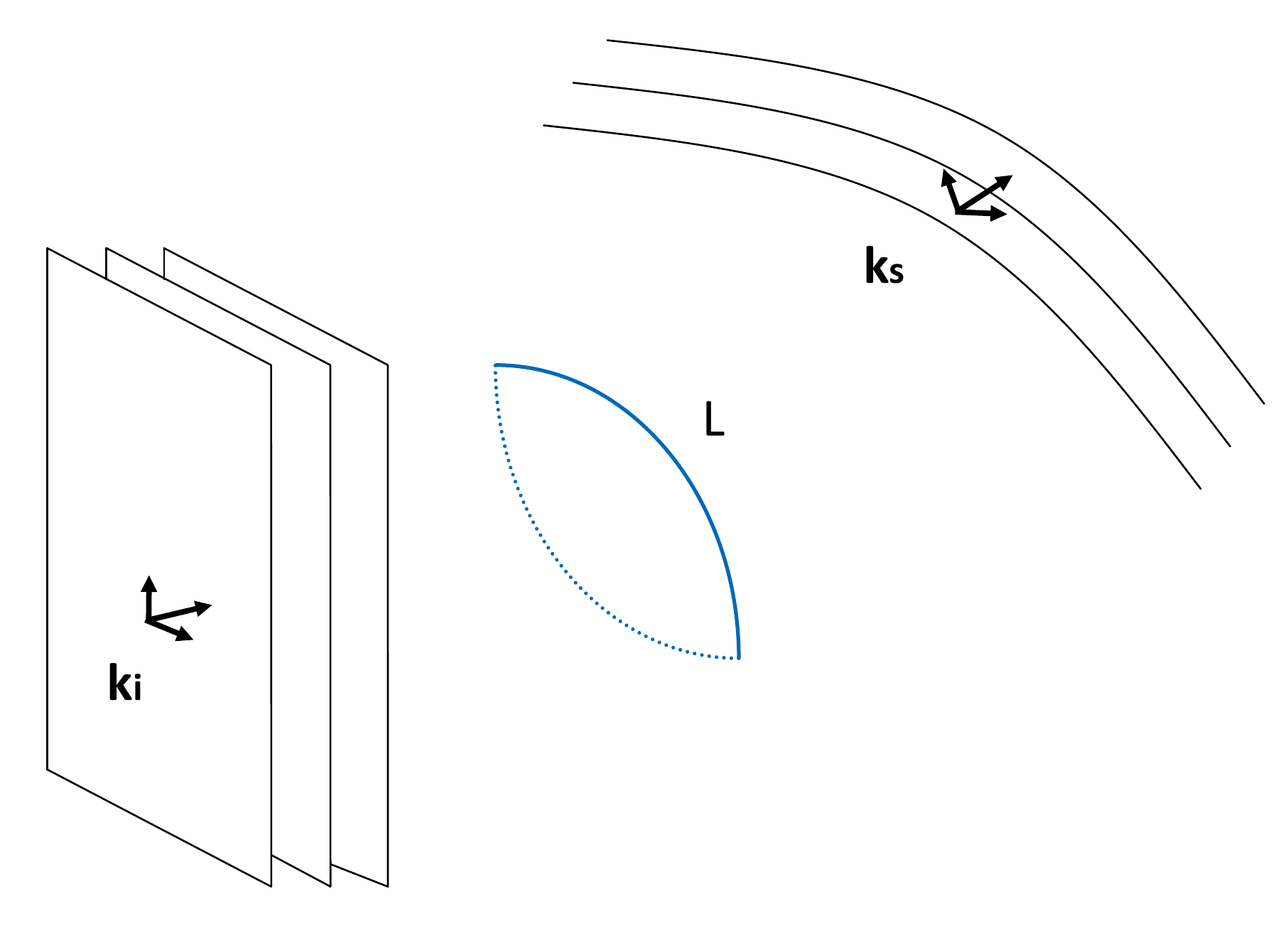}
\caption{The basic scattering mechanism of an elastic wave by one oscillating line dislocation: an  elastic wave of wave vector $\vec k_i$ is incident upon a dislocation segment of length $L$ that oscillates in response. As it does, it re-emits waves with scattered wave vector $\vec k_s$.}
\label{fig_scatt1}
\end{figure}

When a single dislocation is present (see Fig. \ref{fig_scatt1}),  standard scattering theory can be used to turn Eqn. (\ref{EQondes}) into an integral equation that involves the Green function (impulse response function) for the medium under consideration and it can be solved, for example using a Born approximation scheme. This has been done previously  for a whole space\cite{singledisloc} and for a half space with a free surface\cite{surfscatt}.

\subsection{Perturbation approach to multiple scattering}
\begin{figure}[h!]
\includegraphics[width=.8\columnwidth]{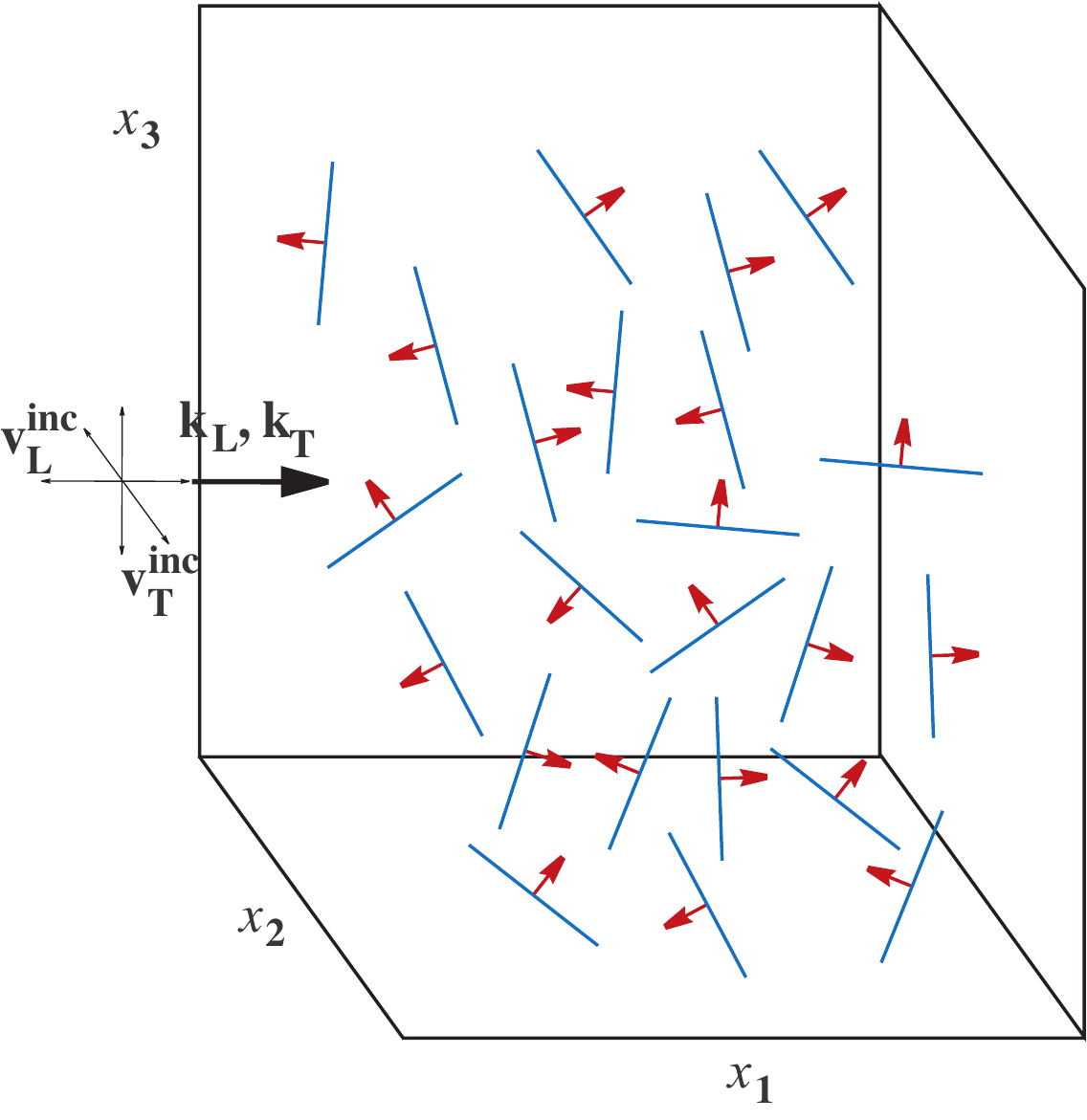}
\caption{Configuration for the study of multiple scattering: the incident wave has transverse (T) and longitudinal (L) polarizations and propagates through an ensemble of randomly distributed and oriented dislocation segments with pinned ends. Each segment is of variable length $L$. Reproduced from Ref. \cite{manydislocs}}
\label{manydislocs}
\end{figure}
When many dislocations randomly distributed are present, as in Figure \ref{manydislocs}, one studies the average (over all possible configurations with prescribed probability distributions) displacement of the elastic medium rather than the displacement itself. The problem to be solved is to find effective plane wave solutions, with a dispersion relation between frequency and effective wave number. Said dispersion relation is obtained from the poles of $\langle G \rangle$ (the brackets denote ensemble averaging) in Fourier space, where $G$ is the impulse response (Green's) function of the dislocation-filled medium, the solution of
\bea
 \rho \omega^2 G_{im}(\vec x,\omega)+c_{ijkl}\frac{\partial^2}{\partial x_j\partial x_l} G_{km}  (\vec x,\omega)=  & & \nonumber \\
 \hspace{1em} - \sum_{\text{disloc. lines}} V_{ik}G_{km}(\vec x,\omega) -
  \delta_{im} \delta(\vec x)  \, . & &
\label{EqOndemodif}
\eea
To this end the average Green's function is written as (``Dyson equation'')
\begin{equation}
 \langle G\rangle ^{-1}= (G^0)^{-1}-\Sigma
\label{dyson}
\end{equation}
where $G^0$ is the Green's function of the medium without dislocations (i.e., ``free''). For an infinite, homogeneous and isotropic medium it is, in Fourier space,
\beq
[(G^0)^{-1}]_{ik}(\vec k,\omega)= (-\rho \omega^2+\mu k^2)\delta_{ik}+(\lambda+\mu)k_ik_k
\label{g0}
\eeq
or
\begin{equation}
{G^0_{ik}}(\vec k,\omega)= \frac{1}{\rho c_T^2(k^2-  k_T^2)}(\delta_{ik}-\hat k_i \hat k_k)+\frac{1}{\rho c_L^2(k^2-k_L^2)} \hat k_i \hat k_k \, .
  \label{g0Sol}
\end{equation} 
with an implied small imaginary part in the denominator to insure causality. It will be explicitely considered later on. { The corresponding expression in coordinate space is defined by 
\beq
G^0_{ik}(\vec x -\vec x') = { \frac{1}{(2\pi)^3} }\int d\vec k e^{-i\vec k \cdot \vec x} {G^0_{ik}}(\vec k) e^{i\vec k \cdot \vec x'} ,
\label{g0}
\eeq
and we shall omit the frequency argument $\omega$ in what follows.} In the independent and weak scattering approximation\cite{ScatRMP} $\Sigma = n\langle T\rangle$, with $T$ the scattering $T$-matrix and $\langle \cdot\rangle$ an average over the internal variables of the dislocation segment: Length, orientation, and Burgers vector. Their position has been assumed to be uniformly distributed with density (number per unit volume) $n$. In \cite{manydislocs} we investigated the properties of elastic waves propagating in this disordered medium and obtained expressions for the effective velocities and attenuation coefficients. These quantities were obtained from $\Sigma= n\langle T\rangle$ at a leading order Born approximation for $T$. We now revisit this series expansion, locate divergent terms, and show that the full Born series for $T$ can be summed.  

\section{Scattering by a single dislocation segment: Born series to all orders}

As usual in scattering theory, we consider Eq.(\ref{EQondes}) and Eq.(\ref{potential}) as an integral equation
\beq
v_{i}(\vec x)=v_{i}^0(\vec x)+\int d\vec x' G^0_{ij}(\vec x-\vec x')[V_{jl}v_{l}](\vec x')
\eeq
with $v_{ik}^0(\vec x)$, for instance, an incident plane wave and $G^0$ the free space Green function whose Fourier transform is given in Eq.(\ref{g0}) and Eq.(\ref{g0Sol}). The $T-$matrix determines the full scattering problem because it satisfies
\beq
v_{i}(\vec x)=v_{i}^0(\vec x)+\int d\vec x' G^0_{ij}(\vec x-\vec x')[T_{jl}v_{l}^0](\vec x') \, .
\eeq
We introduce the definition of the $T$-matrix in momentum space through 
\beq 
T_{ik}(\vec k,\vec k')=\int d\vec x d\vec x' e^{-i\vec k \cdot \vec x}T_{ik}(x,x')e^{i\vec k' \cdot \vec x'}
\eeq 
to obtain
\beq
T_{ik}(\vec k,\vec k')=T^{(1)}_{ik}(\vec k,\vec k')+T^{(2)}_{ik}(\vec k,\vec k')+T^{(3)}_{ik}(\vec k,\vec k')\ldots
\eeq
with the first Born approximation
\bea
T^{(1)}_{ik}(\vec k,\vec k')&=&\int d\vec x e^{-i\vec k \cdot \vec x}V_{ik}(x)e^{i\vec k' \cdot \vec x}
\nonumber \\ 
 & =& -{\mathcal A}\mM_{ij}k_j k'_l \mM_{lk} . 
 \label{TBorn}
\eea
{ T}he second Born approximation is computed in Appendix \ref{appa}. { The result is}
\beq
T^{(2)}_{ik}(\vec k,\vec k')={\mathcal A}^2\mM_{ij}k_j I k'_l \mM_{lk}.
\label{2born}
\eeq 
where
\bea
 \label{eye}
I &\equiv& \frac{1}{ (2\pi)^3} \int d\vec q \, {\vec{q}}\, ^{t}(\mM\cdot G^0(\vec q)\cdot \mM )\vec q  \\ 
& = &\int dq q^4\left(\frac{8\pi}{5}\frac{1}{\rho C_T^2(q^2-k_T^2)}+\frac{16\pi}{15}\frac{1}{\rho C_L^2(q^2-k_L^2)}\right) \nonumber ,
\eea
with ${\vec{q}}\, ^{t}$  the transpose of ${\vec{q}}$. The second line is obtained using the properties of $\mM$:
 \beq
\mM_{mm}=0 \, , \hspace{2em} \mM_{ik}\mM_{ik}=\mM_{ki}\mM_{ik}=2 \, .
\label{tracem}
\eeq


Note that $I$ diverges because it is the integral of a quantity that grows $\sim q^2$ at high wavenumbers. This is a { consequence} of the fact that we use a continuum mechanics approach, in which there is no intrinsic cut-off length. 

The third order Born approximation to the $T$ matrix { is worked out in Appendix \ref{appb}. The result is}:
\beq 
T^{(3)}_{ik}(\vec k,\vec k')   = -{\mathcal A}^3\mM_{ij}k_j I^2 k'_l \mM_{lk}
\label{power}
\eeq
It is apparent from (\ref{power}) that a power series structure suggests itself. To pursue this possibility we consider the $p$-th term in the perturbation series and write it, symbolically, as
\beq
T^{(p)}_{ik} = ( VG^0VG^0 \cdots V )_{ik}
\label{sigmap}
\eeq
where the potential $V$ appears $p$ times. Replacing $G^0$ by its Fourier transform and repeatedly using (\ref{tracem}) as well as the properties of the delta function, leads to
\beq
T^{(p)}_{ik}(\vec k,\vec k')   =  - {\cal A} \mM_{ij}k_j  \left(-{\cal A} I\right)^{p-1} k'_l \mM_{lk} \, ,
\label{powerseries}
\eeq
a power series, as was hoped, and formally we have
\bea
T_{ik}& = & M_{il}k_l\left( \frac{-\mathcal A}{1+{\mathcal A} I} \right) k'_pM_{pk}
\label{TDisloc} \\
& = & T^{(1)}_{ik} \frac{1}{1+{\mathcal A} I} ,
\eea
With $T^{(1)}_{ik}$ given by (\ref{TBorn}). 
From the $T$  matrix it is possible to obtain, following the standard definitions, scattering amplitudes and scattering cross sections. For instance, the longitudinal to longitudinal scattering amplitude is
\beq
f_{LL}(\hat{x})=(\hat{x}\cdot\mM\cdot \hat{x}) \left(\frac{-{\mathcal A}}{1+{\mathcal A} I } \right)(\hat{k_0}\cdot \mM\cdot \hat{k_0})
\label{Longit}
\eeq
where $\hat{x}$ is the scattering direction  and $\hat{k_0}$ the incident direction. Eqn. (\ref{Longit}) is a simple generalization, to all orders, of a similar expression, obtained in I to first order. The only difference is the new denominator. The scattering amplitudes for mode conversion, as well as for transverse to transverse scattering, follow similarly, as does the total scattering cross section using the optical theorem. 

\section{Mass operator}
We return now to the evaluation of the mass operator $\Sigma= n\langle T\rangle$, where the average is over the internal variables of the dislocation segment: Length, orientation, and Burgers vector. Following II, we consider edge dislocations, all with the same length $L$, with a Burgers vector of fixed magnitude but randomly oriented. Dislocation position has been assumed to be uniformly distributed with density (number per unit volume) $n$. Thus
\beq
\Sigma_{ij} = \frac{-n{\cal A} {\cal M}_{ij}}{(1+{\cal A}I)}
\label{sigmatot}
\eeq
where
\beq
{\cal M}_{ij} \equiv \int d\mathbf{C}\mM_{ip}\mM_{qj}k_pk_q
\label{average}
\eeq
and the integration over $\mathbf{C}$ is the average over internal variables. If we take a distribution of dislocations that is isotropic on average we have\cite{manydislocs}
\bea
\langle {\cal M}_{ij} \rangle & = & \frac{1}{15} ( \hat k_i \hat k_j  + 3\delta_{ij}) k^2 \nonumber \\
 & = & \frac 15 ( \delta_{ij} - \hat k_i \hat k_j ) k^2 + \frac{4}{15}  k_i  k_j
\eea
where the brackets denote  angular average. There are two objects of interest: One is the mass operator (\ref{sigmatot}) itself. Without the density factor $n$ and without the average indicated in (\ref{average}) it corresponds to the scattering amplitude by a single dislocation segment, the $T$-matrix. Another is the Dyson equation (\ref{dyson}), where the mass operator appears, and which provides, through the vanishing of a determinant,  an implicit relation between frequency and wavenumber, from which (dispersive) effective phase velocities and attenuations can be read.

Consider then the quantity 
\beq
\sigma^0 \equiv  \frac{\cal A}{1+{\cal A}I} \, .
\label{sigmaDef}
\eeq
that appears in the mass operator (\ref{sigmatot}). The object of interest is the quantity $I$ defined by (\ref{eye}). Introducing the change of variables
\begin{eqnarray}\label{eq53}
\Omega_{T} & = & c_{T}p\nonumber\\
\Omega_{L} & = & c_{L}p \, ,
\end{eqnarray}
and using
\beq
\lim_{\eta \rightarrow 0} \frac{1}{x+i\eta} = {\cal P} \frac 1x + i\pi \delta (x)
\eeq
where ${\cal P}$ means Cauchy principal value, 
one has
 \begin{eqnarray}
I & = & \frac{1}{(2\pi)^{3}\rho c_{T}^{5}} \frac{24\pi}{15} \int \frac{d\Omega_{T} \Omega_{T}^{4}}{\Omega_{T}^{2}-\omega^{2}-\imath \eta} \nonumber \\
& & \hspace{1em} +\frac{1}{(2\pi)^{3}\rho c_{L}^{5}} \frac{16\pi}{15} \int \frac{d\Omega_{L} \Omega_{L}^{4}}{\Omega_{L}^{2}-\omega^{2}-\imath \eta} \nonumber \\
& = & \frac{1}{15\pi^{2}}\left[\frac{3\gamma^{5}+2}{\gamma^{5}}\right]\frac{1}{\rho c_{T}^{5}}\int \frac{d\Omega \Omega^{4}}{\Omega^{2}-\omega^{2}-\imath \eta}  \\
 & \equiv & I_{\cal R} +i I_{\cal I}   \nonumber
\label{eye1}
\end{eqnarray}
with
\bea
I_{\cal R} & = & I_0 \Re \int \frac{d\Omega \Omega^{4}}{\Omega^{2}-\omega^{2}-\imath \eta} = I_0 {\cal P} \int \frac{d\Omega \Omega^{4}}{\Omega^{2}-\omega^{2}}
\label{diverge} \\
 I_{\cal I} & = & I_0 \Im \int \frac{d\Omega \Omega^{4}}{\Omega^{2}-\omega^{2}-\imath \eta}= I_0 \frac{\omega^{3}\pi}{2}
\label{eq63}
\eea
and
\beq
I_0 \equiv \frac{1}{15\pi^{2}}\left[\frac{3\gamma^{5}+2}{\gamma^{5}}\right]\frac{1}{\rho c_{T}^{5}} \, .
\eeq
We see that $I_{\cal R}$, the real part of $I$, diverges since its integrand grows as $\Omega^2$ at large $\Omega$. The imaginary part $ I_{\cal I}$ is finite.  Consider the divergent quantity  $I_{\cal R}$, given by (\ref{diverge}).  Let 
\beq
I_{\cal R} =  \lim\limits_{\Lambda \rightarrow\infty} I_{\cal R}(\omega,\Lambda)
\eeq
with
\beq
I_{\cal R}(\omega,\Lambda)= I_0 {\cal P} \int_0^{\Lambda} \frac{d\Omega \Omega^{4}}{\Omega^{2}-\omega^{2}} .
\eeq
This integral can be calculated; it is
\begin{equation}
I_{\cal R}(\omega,\Lambda)=I_0\left( \Lambda \omega^2+\frac{\Lambda^3}{3}-\omega^3 {\rm Arctanh}\frac{\omega}{\Lambda}\right)
\label{primitiva}
\end{equation}
for $\Lambda > \omega$.

Calling $I_{\cal R} (\omega_1) \equiv I_{{\cal R} 1}$ this leads{ , in the limit $\Lambda \to \infty$,} to
\beq
\frac{I_{{\cal R} 1} - I_{{\cal R} 2}}{\omega_1^2 - \omega_2^2} - \frac{I_{{\cal R} 2} - I_{{\cal R} 3}}{\omega_2^2 - \omega_3^2} =  0  \, .
\label{regular}
\eeq
So the difference between two weighted differences of $I_{\cal R}$, as specified by (\ref{regular}), is finite, indeed it is zero, although $I_{\cal R}$ diverges. It is not unusual in physics to have a finite effect appear as the difference of two infinite quantities; the Casimir effect\cite{casimir} is a notorious example. So, it is natural to interpret (\ref{regular}) as saying that it provides the value of $I_{{\cal R} }$ at a frequency $\omega_3$ in terms of the value of $I_{{\cal R}}$ at two other frequencies $\omega_1$ and $\omega_2$. These two values must be independently established.

The form of the interaction in Eqns. (\ref{EQondes}-\ref{potential}) suggests that the Born series is a  long wavelengths,  or low frequency, approximation. Indeed, since the interaction potential as two gradients, high wave numbers will be more strongly affected than low wave numbers. We should then have
\beq
 \lim_{\omega \to 0} I(\omega) = 0
\eeq
a condition that is already satisfied by the imaginary part $ I_{\cal I}$. So, suppose we wish to have $ I_{\cal R} (\omega)$ at a finite frequency; consider $\omega_1, \omega_2$ small enough that $I$ vanishes. Eqn. (\ref{regular}) provides
\beq
I_{\cal R} (\omega) = 0 \, ,
\label{Realzero}
\eeq
in the sense that $I_{\cal R} (\omega)$ is negligible with respect to $I_{\cal I} (\omega)$ at low frequencies. This last point will
be elucidated in the next section, by a proper scaling through the introduction of cut-off regularization (Eq. \ref{sigma}). Then, from Eqs.(\ref{sigmaDef},\ref{Realzero}) we have, at low frequencies,
\beq
\sigma^0_{\rm low \hspace{.2em} \omega} =  \frac{\cal A}{1+i{\cal A}I_{{\cal I}}} \, .
\label{lowfreq}
\eeq
This is an expression that provides a (finite) correction, valid at low frequencies, to the first order Born approximation $\sigma^0_{\rm Born} =  {\cal A}$.
In order to be more precise about what we mean by ``low frequencies'' we now introduce a cut-off dependent regularization.

\subsection{Introduction of a high frequency cut-off}
The reason $I_{\cal R}$, given by (\ref{diverge}) diverges is that it considers all frequencies, up to infinitely high frequencies, which means all wavelengths, down to infinitely short wavelengths. This is an artifact of the perturbation theory approach, in which each term in the perturbation series introduces an integration over an infinite frequency interval.  More precisely, the order $n$ term has $n-1$ integrals that extend down to very short wavelengths. It has already been shown, in Eqn. (\ref{powerseries}), that these $n-1$ integrals are the $n-1$ power of a single integral $J$. The infinite extent of the integration range, in turn, is a consequence of using continuum mechanics that has no intrinsic length scale, as opposed to real solids that have at least one microscopic length scale given by the interatomic distance. In the problem we are considering in this paper, there is an additional length scale given by the length $L$ of the dislocation segments, and the potential (\ref{potential}) has been obtained for wavelengths long compared to $L$. Accordingly, we now   attempt replacing $I_{\cal R}$ by the finite, but cut-off dependent, quantity $I_{\cal R}(\omega,\Lambda)$ given by (\ref{primitiva}). How is $\Lambda$ to be determined?
We can not fix a precise value for the cutoff $\Lambda$, but we can provide and estimate by noticing that the divergence of $I$ is controlled  by the behaviour of $G^0(q)$ for $|q|\to\infty$. 
Now, $G^0(q)$ appears in the expression for the $T$-matrix because we use Eq.(\ref{potential}) for the potential. This is an approximate expression valid for waves of wavelength longer than the dislocation segment\cite{singledisloc,surfscatt}. For short wavelengths (large wave numbers) the dislocation dynamics can not be described by the exciting field in a reference position [for instance the middle of the dislocation segment as we take for long wavelength to get Eq.(\ref{potential})]. Accordingly for large wavenumbers the potential is different 
and that will change the behaviour of the integrand in $I$  for $|q|\to\infty$. Therefore we chose an upper cut-off $\Lambda=d\omega_1$ that is proportional to the first resonance frequency $\omega_1=\sqrt{\Gamma/m}(\pi/L)$, with constant of proportionality $d$.   Eq.(\ref{potential}) is expected to be valid up to these frequencies for a real material with dislocations of length $L$. 

Consider then $\sigma^0$, given by (\ref{sigmaDef}), with $I$ evaluated using the cutoff $\Lambda=d\omega_1$, ($0<d<1$). Using (\ref{lowS}) and  $ {\rm Arctanh}\frac{\omega}{\Lambda}\approx \frac{\omega}{\Lambda}$ we get 
\beq
\sigma^0 = \frac{aL}{a_1\omega^2-a_2\omega_1^2+i\frac{B\omega}{m}+aLI_0\left( i\frac{\pi}{2}\omega^3-\frac{\omega^4}{d\omega_1}\right)} 
\label{sigma}
\eeq
with 
\bea
a& = & \frac{8}{\pi^2} \frac{\mu^2 b^2}{m} \\
a_1&=&1+aLI_0d\omega_1\\
a_2&=&1-aLI_0\frac{d^3}{3}\omega_1
\eea
It is a straightforward exercise to verify that (\ref{sigma}) leads to  (\ref{lowfreq}) in the limit $\omega\ll d\omega_1\ll \omega_1$.  

\section{Comparison with the renormalization of a Dirac delta potential in a Schr\"odinger equation}
\label{ApB}
The algebra of the problem we are dealing with has some similarities with the scattering of quantum (de Broglie) waves by an isotropic point obstacle as described by a Schr\"odinger equation in $D$ dimensions:
\beq
\left( \nabla^2 + k^2 \right) \psi (\vec x) = g \delta^{(D)} (\vec x) . 
\label{schro}
\eeq
This problem, because of its simplicity, has received some attention in the literature and it is instructive to compare it with our own. The interaction term in Eqn. (\ref{schro}) as our own Eqn. (\ref{potential}), is localized at a point. The main differences between this Schr\"odinger problem and our dislocation problem are: a) the former is scalar and the latter vectorial and b) the potential in the dislocation case Eq.(\ref{potential}) has two gradient operators. In the Schr\"odinger problem $\langle k|V|k'\rangle=g/(2\pi)^D$ i.e. a constant. In the dislocation problem $\langle k|V|k'\rangle=-{\mathcal A}({\mM}\vec k\cdot {\mM}\vec k')$ as shown in Eq.(\ref{TBorn}).  Note also that the two dimensional problem of transverse (anti-plane) waves scattered by an infinite straight screw dislocation is a scalar Schr\"odinger problem\cite{JF} with a potential $\langle k|V|k'\rangle$ proportional to $k^2$. The analysis we review below also holds when the point obstacle is modelled by even derivatives of the delta function\cite{renorm2}. 

When $D=1$ the scattering problem associated to Eq.(\ref{schro}) poses no difficulty. When $D=2$ the scattered cylindrical wave (and/or $G^{0+}$) diverges logarithmically at the origin thus creating a divergence in the $T-$matrix. Similarly, when $D\ge3$ the scattered spherical wave (and/or $G^{0+}$) diverges as $1/r^{D-1}$ creating also a divergence in $T$.  The idea that has been put forward to deal with this divergence is that the potential strength $g$ in $V(\vec x)=g\delta^{(D)}(\vec x)$ is not a measurable quantity but the scattering cross section $\langle k|T|k\rangle$ and possible bound state energies are\cite{renorm1}: In terms of the potential and the Green function one has
\[
\langle  k|T|k'\rangle=\langle  k|V|k'\rangle+\langle  k|VG^0V|k'\rangle+\langle  k|VG^0VG^0V|k'\rangle-...
\]
a quantity we already computed for the dislocation problem. Analogously, for the Schr\"odinger problem we find
\begin{equation}
\langle  k|T|k'\rangle= \left(\frac{1}{2\pi} \right)^D \left(\frac{1}{g}+ J'(k)  \right)^{-1}
\label{schroT}
\end{equation}
 with 
 \[
 J'(k)=\int \frac{d^D q}{(2\pi)^D}\frac{1}{q^2-k^2-i\eta}
 \]
Since $J'(k)$ diverges for $D\geq 2$ one introduces a regulator $\Lambda$ that cuts off the integral for $|q|>\Lambda$
and the previous equation is interpreted as
 \[
\langle  k|T|k'\rangle=\left(\frac{1}{2\pi} \right)^D \lim_{\Lambda\to\infty}\left(\frac{1}{g(\Lambda)}+ J'(\Lambda,k)  \right)^{-1}
\]
in which the unmeasurable parameter in the strength of the potential is turned into a function of $\Lambda$:  $g=g(\Lambda)$. Its form is then determined through the requirement that the scattering matrix be finite for some wave vector $k=k_0$, because $T(k_0)$ is an observable while $g$ is not. Thus
\[
g(\Lambda)\equiv g(T(k_0),J'(\Lambda,k_0)).
\]
 
 To be explicit let us consider $D=5$  
 \bea
 \label{jotap}
 J'(\Lambda,k) & = & \frac{\Omega_5}{(2\pi)^5}\int_0^\Lambda \frac{q^4dq}{q^2-k^2-i\eta} \\
 & =& \frac{\Omega_5}{(2\pi)^5}\left( \Lambda k^2+\frac{\Lambda^3}{3}-k^3 {\rm Arctanh}\frac{k}{\Lambda}+i\frac{k^3\pi}{2}\right) \nonumber
 \eea
where $\Omega_5$ is the solid angle in $D=5$ dimension. Note the similarity between the integral in $J'$ with $I$ from Eq.(\ref{eye1}) in our own dislocation problem.

Since it is not possible to cancel in Eq.(\ref{schroT}) the different diverging functions of $\Lambda$ that appear in Eq.(\ref{jotap}) with $1/g(\Lambda)$, because the former contains $k$ dependent diverging terms and the latter not, one finds from Eq.(\ref{schroT}) that $\lim_{\Lambda\to\infty}\langle  k|T|k'\rangle=0$. In other words, there is no scattering by a delta function potential when the interaction is governed by the Schr\"odinger equation. This is the case for all $D\geq 4$. 
The cases $D=2$ and $D=3$ are different. For instance with $D=3$
 \bea
 J'(\Lambda,k)&=&\frac{\Omega_3}{(2\pi)^3}\int_0^\Lambda \frac{q^2dq}{q^2-k^2-i\eta}\nonumber\\
 &=&\frac{1}{2\pi^2}\left(\Lambda-k{\rm Arctanh}\frac{k}{\Lambda}+ik\frac{\pi}{2}\right)\nonumber
 \eea
and if we measure the $T$-matrix at some $k_0$ we get
\[
\lim_{\Lambda\to \infty}\frac{1}{g(\Lambda)}=[(2\pi)^3T(k_0)]^{-1}-\lim_{\Lambda\to \infty}J'(\Lambda,k_0)
\]
thus it is posible to define a renormalised coupling parameter $g_{\rm RN}$ such that $1/g(\Lambda)=1/g_{\rm RN}-\frac{\Lambda}{2\pi^2}$, i.e. the divergence is absorbed in the unmeasurable coupling parameter, which gives in Eq.(\ref{schroT}) as $\Lambda\to\infty$ a finite value for the $T$-matrix
 \[
\langle  k|T|k'\rangle=\left(\frac{1}{2\pi}\right)^3\left(\frac{1}{g_{\rm RN}}-ik\frac{1}{4\pi}\right)
\]
with $g_{\rm RN}$ a parameter to be determined by a measurement at some given $k_0$. Similarly in $D=2$ 
the diverging term of $J'$ can be absorbed in the coupling parameter $g(\Lambda)$.  
Moreover in the case of an attractive delta potential $g_{\rm RN}$ can be expressed in terms of the energy of the bound state (which is measured).  In the case of repulsive delta potential the $T$-matrix (scattering cross section) must be measured at some value of $k$ and the computed expression gives the value on the rest of the $k$ axis. 

Let us now go back to the dislocation scattering problem. Given the similarity between (\ref{primitiva}) and (\ref{jotap}), one is tempted to perform the renormalisation procedure that will lead to the conclusion that there is no scattering, as in the Schr\"odinger case for $D=5$. But the key difference is that  
the potential for the dislocation involves measurable quantities such as elastic constants, frequency and Burgers vector.
Thus we conclude that the finiteness of the cutoff value is unavoidable in this context.   

Let us finally remark that in the scattering problem of an anti-plane elastic wave by a screw dislocation in two dimensions the above  analogy with the scattering of
a Dirac delta holds, with  $D=4$.

\section{Discussion}

The full Born series for the scattering amplitudes by a single edge dislocation segment as described by the potential Eq.(\ref{potential}) is a geometric series that can be summed and expressed analytically. We have shown that apart form a multiplicative factor, the angular dependence of these scattering amplitudes are the same as in the leading order Born approximation. The multiplicative factor is a frequency and cutoff dependent quantity, that diverges if the cutoff goes to infinity. This divergence is an artifact of the perturbation series within a context of continuum mechanics, where there is no intrinsic, short-distance, length scale. At low frequencies, it has been shown that the seemingly divergent integral can be regularized to zero. As frequencies increase, the integral can be regularized through the introduction of a cut-off. frequency, proportional to the first resonant frequency of the dislocation segment. 

\begin{widetext}
From an experimental point of view we can ask how the cutoff $\Lambda$ can be determined. For this we note that from Eq.(\ref{primitiva}) in the large $\Lambda$ limit (where ${\rm ArcTanh}(\epsilon)\approx \epsilon$) we can get
\[
\frac{I_{\cal R}(\omega_1,\Lambda)-I_{\cal R}(\omega_2,\Lambda)}{\omega_1^2-\omega_2^2}-\frac{I_{\cal R}(\omega_2,\Lambda)-I_{\cal R}(\omega_3,\Lambda)}{\omega_3^2-\omega_3^2}=\frac{1}{\Lambda}\left(\frac{\omega_2^4 -\omega_3^4}{\omega_2^2-\omega_3^2}-\frac{\omega_1^4-\omega_2^4}{\omega_1^2-\omega_2^2}\right) \, .
\]
Thus, weighted differences of $I$ at different frequencies can determine $\Lambda$. The consequence then appears to be that measuring the cross section at two different low frequencies will provide the appropriate value of $\Lambda$ which is crucial to predict the cross section at any, third, frequency. And similarly for the mass operator. It appears then that in order to have a relation between frequency and wave vector for the coherent wave two parameters are needed, to be determined through measurement of $\omega(\vec k)$ at two different points.


\acknowledgements
This work was supported by Fondecyt Grant 1130382  and ANR-CONICYT grant 38, PROCOMEDIA. 
{
\appendix
\section{Second order Born approximation}
\label{appa}
The second order Born approximation is given by 
\bea
T^{(2)}_{ik}(\vec k,\vec k')&=&\int d\vec x \, d\vec x' e^{-i\vec k \cdot \vec x}V_{ij}(x)G^0_{jl} (x-x') V_{lk} (x') e^{i\vec k' \cdot \vec x'} \\
& = & \frac{{\cal A}^2}{(2\pi)^3} \int d\vec x \, d\vec x'  d \vec q \, \mM_{in} \mM_{mj} \mM_{lp} \mM_{kq} e^{-i\vec k \cdot \vec x} \frac{\partial}{\partial x_n} \delta (\vec x - \vec X_0) (-iq_m) e^{-i\vec q \cdot \vec X_0} G_{jl}^0(\vec q ) e^{i\vec q \cdot \vec x'} \frac{\partial}{\partial x'_p} \delta (\vec x' - \vec X_0) (ik'_q) e^{i\vec k' \cdot \vec X_0} \nonumber \\
& = & \frac{{\cal A}^2}{(2\pi)^3} \int d\vec q \, \mM_{in} \mM_{mj} \mM_{lp} \mM_{kq} k'_q k_n q_m q_p G_{jl}^0(\vec q ) \nonumber \\
& = & {\cal A}^2 \mM_{in} k_n I \mM_{kq} k'_q  \, . \nonumber
\eea
The second line is obtained using the expressions (\ref{potential}) for the potential $V_{ij}$ and (\ref{g0}) for the bare Green function $G^0_{jl}$. The third line is obtained performing the integrations over $\vec x$ and $\vec x'$ with the aid of the delta functions. Finally, the last line uses the definition (\ref{eye}) of $I$ and gives (\ref{2born}).

\section{Third order Born approximation}
\label{appb}
The third order Born approximation is given by
\bea
T^{(3)}_{ik}(\vec k,\vec k')   & = &  \int d\vec x \, d\vec x' d\vec x'' \; e^{-i\vec k \cdot \vec x} V_{in}(\vec x)  G^0_{nl}(\vec x-\vec x') V_{lj}(\vec x') G^0_{jm}(\vec x'-\vec x'')V_{mk}(\vec x'') e^{i\vec k' \cdot \vec x''} \, .
\label{third term}
\eea
Proceeding in the same way as for the second order gives Eqn. (\ref{power}):
\[
T^{(3)}_{ik}(\vec k,\vec k')   = -{\mathcal A}^3\mM_{ij}k_j I^2 k'_l \mM_{lk} \, .
\]
}
\end{widetext}

\end{document}